\documentclass{article}
\usepackage[T1]{fontenc}
\usepackage{wrapfig}
\usepackage{dsfont}
\usepackage[portuges,english]{babel}
\usepackage{amssymb,latexsym,amsmath,color,mathrsfs,ifsym,graphics,stmaryrd} 
\usepackage[colorlinks,linkcolor=blue,urlcolor=blue,citecolor=black,
plainpages=false,pdfpagelabels,breaklinks]{hyperref}
\newtheorem{remark}{\sc{Remark}}[section]
\usepackage[all]{xy}

\title{{\sc Relational Quantum Entanglement Beyond\\Non-Separable and Contextual Relativism}}

\author{{\sc Christian de Ronde}$^{1,2,3,4}$ and {\sc C\'esar Massri}$^{5,6}$}
\date{}

\usepackage[margin=2.18 cm]{geometry}

\begin{document}

\bibliographystyle{plain}
\maketitle

\begin{center}
\begin{small}
1. Philosophy Institute Dr. A. Korn, University of Buenos Aires - CONICET\\
2. Institute of Engineering - National University Arturo Jauretche, , Argentina.\\
3. Center Leo Apostel for Interdisciplinary Studies - Vrije Universiteit Brussel, Belgium.\\
4. Federal University of Santa Catarina - Brazil.\\ 
5. Institute of Mathematical Investigations Luis A. Santal\'o, UBA - CONICET\\
6. University CAECE, Argentina.
\end{small}
\end{center}

\bigskip

\begin{abstract}
\noindent In this paper we address the relativist-perspectival nature of the orthodox definition of quantum entanglement in terms of {\it preferred factorizations}. We also consider this aspect aspect within the generalized definition of entanglement proposed by Barnum et al. \cite{BKOV03, BKOV04} in terms of {\it preferred observables}. More specifically, we will discuss the {\it non-separable relativism} implied by the orthodox definition of entanglement, the {\it contextual relativism} implied by its generalization as well as some other serious problems  presently discussed within the specialized literature. In the second part of this work, we address a recently proposed objective-invariant definition of entanglement understood as the actual and potential coding of effective and intensive relations \cite{deRondeMassri19b}. Through the derivation of two theorems we will show explicitly how this new objective definition of entanglement is able to escape both non-separable relativism and contextual relativism. According to these theorems, within this proposed relational definition, all possible subsets of observables as well as all possible factorizations can be globally considered as making reference to the same (potential) state of affairs. The conclusion is that, unlike with the orthodox definitions, this new objective-relational notion of entanglement is able to bypass relativism right from the start opening the door to a realist understanding of quantum correlations.  
\end{abstract}

\renewenvironment{enumerate}{\begin{list}{}{\rm \labelwidth 0mm
\leftmargin 0mm}} {\end{list}}

\newcommand{\ita}{\textit}
\newcommand{\mcal}{\mathcal}
\newcommand{\mfrak}{\mathfrak}
\newcommand{\mbb}{\mathbb}
\newcommand{\mrm}{\mathrm}
\newcommand{\msf}{\mathsf}
\newcommand{\mscr}{\mathscr}
\newcommand{\lra}{\leftrightarrow}
\renewenvironment{enumerate}{\begin{list}{}{\rm \labelwidth 0mm
\leftmargin 5mm}} {\end{list}}

\newtheorem{dfn}{\sc{Definition}}[section]
\newtheorem{thm}{\sc{Theorem}}[section]
\newtheorem{lem}{\sc{Lemma}}[section]
\newtheorem{cor}[thm]{\sc{Corollary}}
\newtheorem{definition}{\sc{Definition}}[section]
\newtheorem{theorem}{\sc{Theorem}}[section]
\newtheorem{corollary}{\sc{Corollary}}[section]
\newcommand{\Proof}{\textit{Proof:} \,}
\newcommand{\cqd}{\hfill{\rule{.70ex}{2ex}} \medskip}

\bigskip

\bigskip

\bigskip

\bigskip

\bigskip

\bigskip

\section{The Strange (Hi)Story of Quantum Entanglement}

The notion of entanglement ({\it Verschr\"ankung}) was introduced in Quantum Mechanics (QM) by Albert Einstein, Boris Podolsky and Nathan Rosen in their famous 1935 ``EPR paper'' \cite{EPR}. That same year, Erwin Sch\"odinger explicitly defined and named the new born concept  \cite{Schr35a}. The main purpose of the introduction of entanglement by Einstein and Schr\"odinger was to expose the inconsistencies present in what had already become the orthodox ``collapse'' interpretation of QM.\footnote{The paradox enlightened by EPR had been conceived by Einstein some years before in discussion with Schr\"odinger. This earlier {\it Gedankenexperiment}, which Einstein regarded as more clear, consisted in entangled particles encapsulated in two different boxes and was finally published in 1948 \cite{Einstein48}.} Both Einstein and Schr\"odinger were trying to expose why ``quantum jumps'' precluded the possibility of the theory of quanta to refer to an objectively represented state of affairs. Something, they feared, had become regarded by the community of physicists as a sort of chimera. As remarked by Alisa Bokulich and Gregg Jaeger, ``[in the EPR paper] the possibility of such a phenomenon [of entanglement]  in quantum mechanics was taken to be a {\it reductio ad absurdum} showing that there is a fundamental flaw with the theory.'' In one of his papers from 1935 Sch\"odinger also remarked an astonishing consequence of the orthodox collapse interpretation of QM, namely, that when two systems get entangled through a known interaction, the knowledge we have of the parts might anyhow decrease.  
\begin{quotation}
\noindent {\small ``If two separated bodies, each by itself known maximally, enter a situation in which they influence each other, and separate again, then there occurs regularly that which I have just called {\it entanglement} of our knowledge of the two bodies. The combined expectation-catalog consists initially of a logical sum of the individual catalogs; during the process it develops causally in accord with known law (there is no question of measurement here). The knowledge remains maximal, but at the end, if the two bodies have again separated, it is not again split into a logical sum of knowledges about the individual bodies. What still remains {\it of that} may have become less than maximal, even very strongly so.---One notes the great difference over against the classical model theory, where of course from known initial states and with known interaction the individual states would be exactly known.'' \cite[p. 161]{Schr35a}}\end{quotation}
Einstein's and Schr\"odinger's criticisms to QM through the notion of entanglement were in fact targeting directly the artificial ``collapse'' of the quantum wave function. A new ``quantum jump'' that had been introduced by Dirac just a few years before in order to bridge the gap between quantum superpositions and single measurement outcomes \cite{Dirac74}. Similar in nature to the original ``quantum jumps'' that Bohr had introduced in his famous 1913 model of the atom, this new {\it ad hoc} process was also completely irrepresentable within the theory. Einstein, was clearly uncomfortable with this unjustified addition. Targeting the collapse, he is quoted by Hugh Everett \cite[p. 88]{Freire15} to have said that he ``could not believe that  a mouse could bring about drastic changes in the universe simply by looking at it''. He also shared his fears with Wolfgang Pauli who would later recall: 
\begin{quotation}
\noindent {\small ``{\it {\small Einstein}}'s opposition to [QM] is again reflected in his papers which he published, at first in collaboration with {\small \emph{Rosen}} and {\small \emph{Podolsky}}, and later alone, as a critique of the concept of reality in quantum mechanics. We often discussed these questions together, and I invariably profited very greatly even when I could not agree with {\small \emph{Einstein}}'s view. `Physics is after all the description of reality' he said to me, continuing, with a sarcastic glance in my direction `or should I perhaps say physics is the description of what one merely imagines?' This question clearly shows {\small \emph{Einstein}}'s concern that the objective character of physics might be lost through a theory of the type of quantum mechanics, in that as a consequence of a wider conception of the objectivity of an explanation of nature the difference between physical reality and dream or hallucination might become blurred.''
\cite[p. 122]{Pauli94}}
\end{quotation}
Schr\"odinger was also disgusted by the introduction of this ``jump'' to which he would refer with great sarcasm:
\begin{quotation}
\noindent {\small``But jokes apart, I shall not waste the time by tritely ridiculing the attitude that the state-vector (or wave function) undergoes an abrupt change, when `I' choose to inspect a registering tape. (Another person does not inspect it, hence for him no change occurs.) The orthodox school wards off such insulting smiles by calling us to order: would we at last take notice of the fact that according to them the wave function does not indicate the state of the physical object but its relation to the subject; this relation depends on the knowledge the subject has acquired, which may differ for different subjects, and so must the wave function.'' \cite[p. 95]{Freire15}} \end{quotation}

Regardless of the strong and clear arguments presented by the two rebels, mainly due to the triumph of Bohr's anti-realist interpretation of QM sealed later by the post-war coming into power of U.S. instrumentalism, the notion of quantum entanglement remained buried for almost half a century. As Jeffrey Bub \cite{Bub17} explains: ``Most physicists attributed the puzzling features of entangled quantum states to Einstein's inappropriate `detached observer' view of physical theory, and regarded Bohr's reply to the EPR argument (Bohr, 1935) as vindicating the Copenhagen interpretation.'' Physicists were not taught about entanglement in Universities and the notion became considered as an anomaly arising from old metaphysical discussions, a ghost which had been silently forgotten by more down-to-earth and useful physical applications.\footnote{As recalled by Lee Smolin \cite[p. 312]{Smolin07}: ``When I learned physics in the 1970s, it was almost as if we were being taught to look down on people who thought about foundational problems. When we asked about the foundational issues in quantum theory, we were told that no one fully understood them but that concern with them was no longer part of science. The job was to take quantum mechanics as given and apply it to new problems. The spirit was pragmatic; `Shut up and calculate' was the mantra. People who couldn't let go of their misgivings over the meaning of quantum theory were regarded as losers who couldn't do the work.''} This was at least until during the first years of the 1980s Alain Aspect and his group in Orsay  were able to experimentally test in the lab the strange correlations exposed in the old EPR thought-experiment \cite{AGR81}. Since it was not possible to measure the correlation between single outcomes ---as originally presupposed in the EPR paper---, Aspect used a reconfigured experiment proposed by John Bell who a few decades before, during his spare time from his real work at C.E.R.N. in particle physics, derived an inequality for the correlations imposed by classical probability ---something that Boole had already done one century before.\footnote{The relation between the famous Bell's inequalities from 1964 and Boole's less known inequalities from 1862, derived almost exactly one century before, has been explicitly considered by Itamar Pitowsky in \cite{Pitowsky94}.} The existence of a violation in the Boole-Bell inequality within an EPR type-experiment implied a violation of classical correlations. This only meant that classical correlations could not represent consistently the data found within the experiment. Period. It did not logically imply ---as it is still presupposed in great part of the physical and philosophical literature---, the existence of entanglement or quantum correlations. Something that should have been obvious to everyone from the fact that Boole-Bell inequalities are a statistical statement about classical probability, not about QM. 

Aspect's experiment became a huge boost for the young discipline called ``philosophy of QM''. But its influence was not restricted to philosophy. The post-war generation of physicists which had been trained with instrumentalist values and a purely pragmatic understanding of physics suddenly became aware of the fact that quantum correlations were much stronger than classical ones. It became evident ---to some of them--- that the existence of entanglement in quantum correlations could be used as a resource of information transfer. As described by Bub \cite{Bub17}, ``[...] it was not until the 1980s that physicists, computer scientists, and cryptographers began to regard the non-local correlations of entangled quantum states as a new kind of non-classical resource that could be exploited, rather than an embarrassment to be explained away.'' In this new context, full of new technical possibilities, the ``shut up and calculate!'' widespread instrumentalist attitude of physicists had to unwillingly allow the reopening of foundational and philosophical debates about QM. Very soon, even physicists would come to regard entanglement ---like Schr\"odinger had done many decades before--- as the most essential feature of QM itself. During the 1990s, even a new sub-discipline was created in order to discuss the phenomena of entanglement, namely, {\it Foundations of Quantum Mechanics}. As part of it, {\it Quantum Information} would also became one of the most important lines of research in physics worldwide. After half a century of hibernation, the problems and notions addressed by Einstein and Schr\"odinger were beginning to be openly discussed not only by philosophers but also by physicists. However, as remarked by Bokulich and Jaeger \cite[p. xi]{BokulichJaeger10} in the introduction to their book, {\it Philosophy of Quantum Information and Entanglement}, ``[During the last decades] a large body of literature has emerged in physics, revealing many new dimensions to our concepts of entanglement and non-locality, particularly in relation to information. Regrettably, however, only a few philosophers have followed these more recent developments, and many philosophical discussions still end with Bell's work.'' One decade after the publication of this book, the situation does not seem to have changed and ---apart from few exceptions--- philosophers of physics have not engaged ---as they certainly could--- in this ongoing technological revolution that is taking place today. 

In this work we attempt to critically address the orthodox definition of quantum entanglement in terms of non-factorizability, as well as its recent generalization in terms of observables \cite{BKOV03, BKOV04}. While the first definition is relative to a {\it preferred factorization} of the state of the quantum system, the latter definition of entanglement is relative to the choice of a subset of {\it preferred observables}. Regardless of their differences, both of these notions are explicitly non-objective, in the sense that they are unable to provide a global, consistent and invariant representation of a state of affairs. The subject must be necessarily introduced in order to choose between different non-consistent factorizations and observables. After critically discussing these two definitions as well as some of their problems already addressed in the specialized literature we continue to discuss a newly proposed objective definition of quantum entanglement grounded on the actual and potential coding of effective and intensive relations \cite{deRondeMassri19b}. We attempt to show, through the explicitly derivation of two theorems, that within this new relational definition of entanglement, all possible factorizations as well as all possible subsets of observables can be conceived as making reference to the same (potential) state of affairs in a global, consistent and operationally invariant manner.




\section{Orthodox Entanglement and Non-Separable Relativism}

The contemporary definition and interpretation of quantum entanglement is intrinsically linked to the triumph, that took place during the 20th Century, of a purely pragmatic understanding of physics. This victory was made possible, firstly, by the unspoken alliance between Niels Bohr ---maybe the most influential physicist of the century--- and positivists, and secondly, by the post-war convergence of these anti-realist forces into an even more radical anti-realist scheme called ``instrumentalism''. At the beginning of the 1960s Karl Popper, one of the major figures within the field of epistemology and philosophy of science, wrote in {\it Conjectures and Refutations} \cite{Popper63} that the realist account of physics ---the idea that physical theories make reference to {\it physis} (or reality)--- had been finally defeated: ``Today the view of physical science founded by Osiander, Cardinal Bellarmino, and Bishop Berkeley, has won the battle without another shot being fired. Without any further debate over the philosophical issue, without producing any new argument, the {\it instrumentalist} view (as I shall call it) has become an accepted dogma. It may well now be called the `official view' of physical theory since it is accepted by most of our leading theorists of physics (although neither by Einstein nor by Schr\"odinger). And it has become part of the current teaching of physics.'' Popper argued there were two main reasons for having reached this result. The first reason was the impact of Bohr's interpretation of QM in the physics community: ``In 1927 Niels Bohr, one of the greatest thinkers in the field of atomic physics, introduced the so-called {\it principle of complementarity} into atomic physics, which amounted to a `renunciation' of the attempt to interpret atomic theory as a description of anything.''\footnote{It is interesting to notice that pointing his finger to Bohr, the neo-Kantian, Popper seems to have overlooked the deep influence of his own tradition within the instrumentalist path. In fact, logical and empirical positivism ---just like Bohr--- had also shifted the center of analysis from the theory to the subject, from representation to observation and from explanation to prediction.} The second reason mentioned by Popper was related to ``the spectacular practical success of [quantum mechanical] applications.'' According to Popper: ``Instead of results due to the principle of complementarity other and more practical results of atomic theory were obtained, some of them with a big bang. No doubt physicists were perfectly right in interpreting these successful applications as corroborating their theories. But strangely enough they took them as confirming the instrumentalist creed.'' Of course, one could also argue in favor of instrumentalism that what physicists had seemingly proved ---specially in the influential Manhattan project--- was that spectacular instrumental applications ---such as the atomic bomb--- could be actually developed without entering philosophical debates about the meaning and the reference of the theory. However, this ---as entanglement would show--- was a very limited conclusion to be drawn. 

As it has been argued in detail in \cite{deRonde20b, deRonde20c}, the anti-realist approach to physics produced an essentially inconsistent discourse, still today firmly in place, grounded in two contradictory claims about the theory of quanta. On the one hand, QM was regarded as a ``tool'' that should be used by agents in order to predict measurement outcomes. As famously argued by Chris Fuchs and Asher Peres: \cite[p. 70]{FuchsPeres00}: ``[...] quantum theory does not describe physical reality. What it does is provide an algorithm for computing probabilities for the macroscopic events (`detector clicks') that are the consequences of experimental interventions. This strict definition of the scope of quantum theory is the only interpretation ever needed, whether by experimenters or theorists.'' But on the other hand, it was also sustained ---even by extreme anti-realists--- that QM made reference to a microscopic realm of unseen and irrepresentable ``elementary particles''. This inconsistent discourse mixing an extreme form of pragmatism together with a fictional microscopic realm was constructed by no one else than Niels Bohr himself. In many different occasions Bohr systematically introduced pictures, principles and concepts which had no link whatsoever to the mathematical formalism of QM nor the experience observed in the lab. A very good example of this procedure is the introduction of ``quantum jumps'' in his own model of the Hydrogen atom. An idea he also applied in his interpretation of Schr\"odinger's wave mechanics. Bohr's line of reasoning was exposed during his meeting in 1926 with Erwin Schr\"odinger in which the existence of ``quantum jumps'' within the atom was a kernel point of disagreement. Under the attentive gaze of Werner Heisenberg, Schr\"odinger presented many arguments exposing not only the lack of any explanatory power of this ``magical process'' but also the contradictions reached when introducing such a-causal ``jumps''. The complete lack of conceptual and theoretical support allowed Schr\"odinger to conclude that ``the whole idea of quantum jumps is sheer fantasy.'' \cite[p. 74]{Heis71} But while the Austrian physicist was unwilling to accept the ``jumps'' ---a critical position that would be shared not only by Einstein but also, as the years passed by, by an elder Heisenberg---, the Danish physicist was ready to blame the lack of representation to the theory of quanta itself. Instead of providing a reply to the many criticism presented by Schr\"odinger, Bohr simply embraced them and turned them completely upside-down as showing the limits of the theory of quanta itself. Like a great Judo Master Bohr didn't try to fight against Schr\"odinger's strong arguments; instead, by inverting the direction of their force, he used them as weapons against his enemy: 
\begin{quotation}
\noindent {\small ``What you say is absolutely correct. But it does not prove that there are no quantum jumps. It only proves that we cannot imagine them, that the representational concepts with which we describe events in daily life and experiments in classical physics are inadequate when it comes to describing quantum jumps. Nor should we be surprised to find it so, seeing that the processes involved are not the objects of direct experience.'' \cite[p. 74]{Heis71}} 
\end{quotation}   
According to Bohr, QM went beyond our classical ``manifest image'' of the world, and thus it came with no surprise that our (classical) concepts were incapable of explaining what was really going on in the quantum world. We can only explain what we see in terms of classical physics because experience is, and will always be, classical. That is the way things are, and there is nothing to be done but accept the limitations we have reached within science, limitations which ---Bohr declared--- are not only technological but also ontological. QM had shown us the limitations in the possibilities of representation of Nature itself. It is in this way that Bohr was able to turn his own incapacity to develop a consistent representation of QM into a proof of the theory's own difficulties and limits. In the second half of the 20th Century this inconsistent discourse was sedimented and widespread within both physics and philosophy communities investigating QM. Ever since, the idea that QM makes reference to ``elementary particles'' has stood as a dogma which, quite regardless of the lack of theoretical and experimental support, cannot be questioned. Bohr also devised an escape route, to be used when pushed to address the theoretical representation of this microscopic realm, which has been learned by both physicists and philosophers. When confronted to such type of questioning about quantum particles you can always declare that ---in fact--- this was ``just a way of talking''.  This inconsistent discourse has had also deep consequences in the constitution of the debates about quantum entanglement. 

During the mid-1990s, the physicist and philosopher Abner Shimony \cite{Shimony95} gave one of the first contemporary definitions of entanglement: ``A quantum state of a many-particle system may be `entangled' in the sense of not being a product of single-particle states.'' Of course, this uncritical application of the notion of ``particle'' is not something that can be exclusively blamed to Shimony. During the 1990s the discussions about entanglement were taking place in an instrumentalist environment with physicists accustomed to talk about ``particles'' in a meaningless fashion, just as ``a way of talking'' ---or, as philosophers preferred to term it, as a ``useful fiction''. It is true that both Einstein and Schr\"odinger used the notion of `particle' in order to discuss about the phenomena of entanglement, but ---let us not forget--- they did so as a {\it reduction ad absurdum}, to show a problem, not a solution.\footnote{In fact, neither Einstein nor Schr\"odinger believed in the need to make reference to `particles'. For example, in a letter to Max Born, Einstein wrote the following: ``We are, to be sure, all of us aware of the situation regarding what will turn out to be the basic  foundational concepts in physics: the point-mass or the particle is surely not among them.''} It is also true that Bell talked about ``particles'' in the 1960s, but in this case he was arguing in favor of finding a classical representation to describe quantum phenomena ---an idea that Einstein found ``to cheap''.\footnote{In a letter to Max Born \cite[p. 44]{Freire15} he commented: ``Have you noticed that Bohm believes (as de Broglie did, by the way, 25 years ago) that he is able to interpret the quantum theory in deterministic terms? That way seems too cheap to me.''} And this possibility ---contrary to Bell's intuition--- was ruled out by Aspect's experiments. Today, the unjustified application of particle metaphysics has reached an unspoken consensus within the specialized literature about  entanglement which can be witnessed in the introduction of most papers on the subject, all of which make explicit reference not only to ``particles'' but also to ``separability'', ``locality'' and ---even--- ``purity''. It is then commonly argued by specialists in the field that entanglement is an ``holistic property of compound quantum systems, which involves nonclassical correlations between subsystems'' \cite[p. 865]{Horodecki09}.  For example, in the introduction to the just mentioned book edited by Bokulich and Jaeger, {\it Philosophy of Quantum Information and Entanglement}, the concept is presented in the following manner: 
\begin{quotation}
\noindent {\small ``Consider two particles, $A$ and $B$, whose (pure) states can be represented by the state vectors $\psi_A$ and $\psi_B$. Instead of representing the state of each particle individually, one can represent the composite two-particle system by another wavefunction, $\Psi_{AB}$. If the two particles are unentangled, then the composite state is just the tensor product of the states of the components: $\Psi_{AB} = \psi_A \otimes \psi_B$; the state is then said to be factorable (or separable). If the particles are entangled, however, then the state of the composite system cannot be written as such a product of a definite state for $A$ and a definite state for $B$. This is how an entangled state is defined for pure states: a state is entangled if and only if it cannot be factored: $\Psi_{AB} \neq \psi_A \otimes \psi_B$.'' \cite[p. xiii]{BokulichJaeger10}} 
\end{quotation}   
The fact that none of these notions ---i.e., particle, separability, locality and purity--- have been  adequately defined ---neither formally nor conceptually--- in the context of quantum theory has remained almost unnoticed to the general physicist community which tends to regard, at least when necessary (i.e., when understanding is required), this same discourse as devoid of reference, ``just as a way of talking'' (see for a detailed discussion \cite{deRondeMassri20b}). This controversial definition and picture of entanglement has not stopped here. It has been also extended to the case of mixed states, another very controversial notion introduced within orthodox textbook QM during the 1940s and 1950s. According to orthodoxy, {\it mixed states} are convex sums of pure states. Extending the non-separability definition of pure states, entangled mixed state are those which cannot be written as a convex combination of products.

$$\rho_{mix} = \sum_{i} p_{i} \  \rho_{i}^{pure} =  \sum_{i} p_i \ | \Psi_i \rangle \langle\Psi_i |$$

\noindent Mixtures imply an ignorance interpretation about the purity of states which, in turn, introduces a distinction between states which are ``purely quantum'' and those which are not, namely, mixtures (of pure states). While the convex elements representing pure states are the ``quantum'' ones, mixtures are understood as being derived from them. In fact, the orthodox understanding of mixtures applies an ignorance interpretation which is forbidden in the general case of quantum probability.  

The introduction of convex sets in order to describe pure states and mixtures has very serious drawbacks and problems which have been addressed in detail in \cite{deRondeMassri19d, deRondeMassri20a}. But even leaving behind these deeper criticisms against the tenability of the whole pure-mixture interpretation of QM, there are serious technical problems coming even from within the orthodox literature about entanglement. One of the main recognized difficulties is the fact that the reference to entanglement is dependent on the chosen factorization of the quantum system. As Alejandro de la Torre et al. have shown explicitly in \cite{DelaTorre10}: ``for any system in a factorizable state, we can find different degrees of freedom that suggest a different factorization of the Hilbert space where the same state becomes entangled.'' Or put more dramatically by Thirring et al. \cite{Thirring11}: ``for a given quantum state, it is our freedom of how to factorize the algebra to which a density matrix refers. Thus we may choose! Via global unitary transformations we can switch from one factorization to the other, where in one factorization the quantum state appears entangled, however, in the other not. Consequently, entanglement or separability of a quantum state depends on our choice of factorizing the algebra of the corresponding density matrix.'' Let us see this in some detail. In their article \cite{DelaTorre10}, the authors prove that given a decomposition of the Hilbert space $\mathcal{H}=\mathcal{H}_A\otimes \mathcal{H}_B$ and a factorizable state $\Psi=\Psi_1\otimes \Psi_2$, there exist transformations $F=A\otimes I+I\otimes B$ and  $G=A\otimes I-I\otimes B$ such that $\mathcal{H}=\mathcal{H}_F\otimes \mathcal{H}_G$ and $\Psi$ is not factorizable. In order to prove this result, the authors use the Quantum Covariant Function \emph{QCF},
\[
QCF(X,Y,\Psi):=\langle \Psi,XY\Psi\rangle - \langle \Psi,X\Psi\rangle\langle \Psi,Y\Psi\rangle.
\]
Notice that a factorizable state $\Psi$ with respect to $A\otimes I$ and $I\otimes B$ satisfies $QCF=0$,
\[
QCF(A\otimes I,I\otimes B,\Psi_1\otimes\Psi_2)=
\langle \Psi_1\otimes\Psi_2,A\otimes B\Psi_1\otimes\Psi_2\rangle - 
\langle \Psi_1,A\Psi_1\rangle\|\Psi_1\|^2 \|\Psi_2\|^2\langle \Psi_2,B\Psi_2\rangle=0.
\]
But the same state $\Psi$ with respect to $F$ and $G$ satisfies $QCF(F,G,\Psi)\neq 0$.
In particular, $\Psi$ is entangled in $\mathcal{H}=\mathcal{H}_F\otimes \mathcal{H}_G$ and this shows mathematically that the notion of entanglement depends on the factorization of the Hilbert space, in other words, it depends on the chosen basis for $\mathcal{H}$. This result has introduced some unease in the foundational literature. According to John Earman, there are two broad reactions to this ``threatened rampant ambiguity'' of quantum entanglement:  
\begin{quotation}
\noindent {\small ``{\it Realism.} The Realist claims that the kind of entanglement we should be concerned with is entanglement over subalgebras that correspond to real as opposed to virtual or fictitious subsystems. Obviously, this distinction cannot be drawn in terms of purely formal conditions on the algebras but must come from background physics. [...] But where the hope is utterly dashed and there is no principled way to draw the distinction, the Realist holds that entanglement ought to be regarded as a moot topic.

\smallskip 

\noindent {\it Pragmatism.} The Pragmatist asserts that there is no metaphysically valid way to draw a line between subsystems that are `real' and those that are `fictitious' or `virtual.' Any decomposition of the system algebra into subalgebras meeting appropriate formal criteria is as good a way as any other for defining subsystems. Thus, contrary to Realism, there is no metaphysically valid distinction to be drawn between `real' entanglement (i.e. entanglement over `real' subsystems) and `faux' entanglement (i.e. entanglement over `fictitious' or `virtual' subsystems).'' \cite{Earman15}}\end{quotation}
This relativist dependence of the notion of entanglement with respect to different factorizations can be resumed in the following manner.

\smallskip 
\smallskip 

\noindent {\it {\sc Non-Separable Relativism:} Given a pure state, the existence of entanglement is relative to the chosen factorization. Consequently, the same state will be entangled or not depending on the factorization considered.}

\smallskip 
\smallskip 

\noindent Due to its relative nature, the orthodox definition of quantum entanglement cannot be interpreted as a property of quantum systems. Instead, it must be regarded as intrinsically {\it relative} to the factorization of systems into subsystems. In this context, entanglement becomes more a way in which the observer chooses to describe systems, than a property of the systems themselves. This idea seems to flagrantly contradict the original understanding of entanglement as a (real) property characterizing the relation between quantum particles, presenting instead a rather controversial reference to the way in which subjects or agents choose to describe the fictitious (or virtual) interaction between quantum systems. Embracing this anti-realist account even more explicitly, a generalized notion of entanglement has been put forward by Barnum, Knill, Ortiz, Somma and Viola in a series of papers to which we now turn our attention.

\section{Generalized Entanglement and Contextual Relativism}

The maturity of perspectival-relativism in QM and its generalized acceptance by the physics community seems to have taken place when Niels Bohr was finally declared the indisputable champion of the EPR battle in 1935. Forced by Einstein into an uncomfortable debate about the definition of the physical reality of entangled particles, Bohr had to state more emphatically than ever his perspectival-relativism with respect to measurement situations and outcomes. Bohr's {\it principle of complementarity}, originally created in order to bring together the inconsistent representation of quantum objects as `waves' and `particles', was now applied to the mathematical formalism of the quantum theory itself in order to ascribe the value of properties as {\it relative} to the choice of the experimental set up. As Bohr \cite[p. 700]{Bohr35} famously argued, the value of complementary properties could not be considered as {\it prior} to the determination of the actual measurement set up since, he argued, ``there is essentially the question of an influence on the very conditions which define the possible types of predictions regarding the future behavior of the system.'' The Danish physicist was applauded by a physics community already in tune with the anti-foundational postmodern {\it Zeitgeist} of the 20th Century in which {\it reality} and {\it truth} were regarded as metaphysical hallucinations. Portrayed as a mystic figure of wisdom and knowledge Bohr was contra posed to Einstein, an old and lonely man who seemed incapable of understanding ---what Bohr claimed to be--- the deep revolutionary consequences of QM. 

Bohr's solution to the EPR paradox was not an {\it objective} one, it was an {\it intersubjective} solution which had the purpose of erasing any reference of the theory of quanta beyond the choice of particular measurement situations and outcomes. Let us be very clear about this point. While `objectivity' meant ---in the terms inherited from Kantian philosophy--- the possibility to represent consistently a {\it moment of unity} in categorical conceptual terms, in the neo-Kantian terms of Bohrians and positivists, `objectivity' was translated either as `faithful observation' or simply as `intersubjectivity'.\footnote{This also becomes explicit in Bohr's understanding of the word {\it phenomena}. While in the Kantian scheme the term {\it phenomena} implied the necessary reference to a conceptual  {\it moment of unity}, for Bohr  `clicks' in detectors and `spots' in photographic plates could be also labeled as {\it phenomena}. As explained by Hilgevoord and Uffink in their excellent entry of the {\it Stanford Encyclopedia} on Heisenberg's relations \cite{HilgevoordUffink01}: ``Central in Bohr's considerations is the language we use in physics. No matter how abstract and subtle the concepts of modern physics may be, they are essentially an extension of our ordinary language and a means to communicate the results of our experiments. These results, obtained under well-defined experimental circumstances, are what Bohr calls the `phenomena'. A phenomenon is `the comprehension of the effects observed under given experimental conditions' (Bohr 1939: 24).''} The essential point is that intersubjectivity did not required the conceptual unity presupposed within objective representations. Bohr had shifted from objectivity to intersubjectivity, but he was not willing to abandon the term `objective'. Thus, he simply renamed `intersubjective statements' as `objective statements'. Stressing the claim that his account of QM was as ``objective'' as classical physics he \cite[p. 98]{D'Espagnat06} argued that: ``The description of atomic phenomena has [...] a perfectly objective character, in the sense that no explicit reference is made to any individual observer and that therefore... no ambiguity is involved in the communication of observation.'' Bernard D'Espagnat explains this quotation in the following manner: ``That Bohr identified objectivity with intersubjectivity is a fact that the quotation above makes crystal clear. In view of this, one cannot fail to be surprised by the large number of his commentators, including competent ones, who merely half-agree on this, and only with ambiguous words. It seem they could not resign themselves to the ominous fact that Bohr was not a realist.'' In Bohr's scheme the experiences acquired by different agents were simply detached from any common {\it objective} reference and representation.\footnote{In his book, \cite{D'Espagnat06}, D'Espagnat clearly distinguishes between {\it objective statements} and Bohr's {\it intersubjective statements},  which he calls: {\it weakly objective statements}.} This also implied a silent replacement of the notion of {\it object} ---a conceptual {\it moment of unity} categorically constituted through the general principles of existence, non-contradiction and identity---; by that of {\it event},  {\it actual observation}, or even, {\it measurement outcome}. These steps were perfectly aligned with Bohr's instrumentalist understanding of the mathematical formalism of QM as making exclusive reference to observable predictions.
\begin{quotation}
\noindent {\small ``The entire formalism is to be considered as a tool for deriving predictions, of definite or statistical character, as regards information obtainable under experimental conditions described in classical terms and specified by means of parameters entering into the algebraic or differential equations of which the matrices or the wave-functions, respectively, are solutions. These symbols themselves, as is indicated already by the use of imaginary numbers, are not susceptible to pictorial interpretation; and even derived real functions like densities and currents are only to be regarded as expressing the probabilities for the occurrence of individual events observable under well-defined experimental conditions.'' \cite[p. 314]{Bohr48} }
\end{quotation}
Bohr stressed repeatedly that the most important lesson we should learn from QM was an epistemological one; namely, that {\it we are not only spectators, but also actors in the great drama of (quantum) existence.} This idea went in line with his pragmatic understanding of physics according to which \cite{Bohr60}: ``Physics is to be regarded not so much as the study of something a priori given, but rather as the development of methods of ordering and surveying human experience. In this respect our task must be to account for such experience in a manner independent of individual subjective judgement and therefore objective in the sense that it can be unambiguously communicated in ordinary human language.'' All these ideas can be condensed in the orthodox widespread claim that in QM ``the properties of a system are different whether you look at them or not'' (e.g., \cite{Butterfield17}). 

When taking for granted ---as Bohr did--- that experience must be necessarily described in terms of classical concepts there seems to exist only two possible ways out, either we must change completely the mathematical formalism (as in Bohm's theory or GRWs proposal), or we must avoid to talk about physical reality beyond measurement outcomes (as QBism has explicitly proposed). This latter anti-realist position to which Bohr contributed so much can be also re-farmed in co-relational terms, something that has been extensively proposed (e.g., \cite{Everett57,Rovelli96}) and has led many \cite{Yang18} to believe that ``a quantum state is relative in nature. That is, an observer independent quantum state is not necessarily the basic description of a quantum system.'' It is this possibility which has been investigated by Howard Barnum et al. in a series of papers \cite{BKOV03, BKOV04, BKOV05} which attempt to generalize the notion of quantum entanglement beyond any objective reference. As recalled by them \cite{BKOV05}: ``The standard definition of quantum entanglement requires a preferred partition of the overall system into subsystems ---that is, mathematically, a factorization of the Hilbert space as a tensor product. Even within quantum mechanics, there are motivations for going beyond such subsystem-based notions of entanglement.'' In order to address quantum entanglement from a more general perspective they propose the following \cite{BKOV05}:  ``the key idea behind GE is that entanglement is an observer-dependent concept, whose properties are determined by the expectations of a distinguished subspace of observables of the system of interest, without reference to a preferred subsystem decomposition.'' Thus, avoiding the reference to different factorizations \cite{ViolaBarnum10}: ``Generalized Entanglement (GE) of a quantum state relative to the distinguished set may then be defined without reference to a decomposition of the overall system into subsystems (15; 16).'' However, GE reintroduces relativism even more explicitly by making reference to a preferred set of observables (or context) \cite{ViolaBarnum10}: ``entanglement is an inherently relative concept, whose essential features may be captured in general in terms of the relationships between different observers''. Let us see how this works in some detail. 

In \cite{BKOV05}, the authors give a definition of entangled state using the notion of cones. Let us start by recalling some of the definitions. A \emph{convex set} $C$ is a set in a real vector space $V$ closed under convex combinations,  that is, if $x,y\in C$ then $tx+(1-t)y\in C$ for all $0\le t\le 1$. An element $x\in C$ is called \emph{extremal} if it cannot be written as a nontrivial convex combination.
In other words, if $x=tx_1+(1-t)x_2$ with $x_1,x_2\in C$ and $0<t<1$, then $x$ is not extremal.
A \emph{positive cone} is a proper subset $K$ of a real vector space $V$ 
closed under multiplication by nonnegative scalars. It is called \emph{regular cone} if it is: 
\begin{itemize}
\item convex (equivalently, closed under addition: $K + K = K$), 
\item generating ($K - K = V$, equivalently $K$ linearly generates $V$) 
\item pointed ($K\cap -K = \{0\}$, so that it contains no non-null subspace of $V$), and 
\item topologically closed (in the norm topology).
\end{itemize}
A \emph{ray} in a cone $K$ is a set given by the intersection of a one-dimensional subspace with $K$. An \emph{extreme ray} $R$ of a cone $K$ is a ray such that no $y\in R$ can be written as a convex combination of elements of $K$ that are not in $R$. The authors show that a cone is convexly generated by its extreme rays. 
\begin{definition}[Def. 3.3 in \cite{BKOV05}]
Let $K_1\subseteq V_1$ and $K_2\subseteq V_2$ be two cones. Let $\pi:V_1\to V_2$ be a linear map such that $\pi(K_1)=K_2$. An element $x\in K_1$ is called \emph{generalized unentangled} if
either (a) $x$ and $\pi(x)$ belong to extreme rays, or (b) $x$ is a positive linear combination (or a limit of such combinations) of elements satisfying (a). Then, an element $x$ is called \emph{generalized entangled} if it is not \emph{generalized unentangled}. Both notions are relative to $K_2$ and $\pi$.
\end{definition}
Let us show that the standard definition of entangled state is a particular case of a generalized entangled state. Let $\mathcal{H}$ be a Hilbert space and consider the real vector space of Hermitian operators $\mathcal{A}$. Inside $\mathcal{A}$, we have the cone $Pos$ of positive operators. The extreme rays of $Pos$ are in bijection with pure states (rank one density matrices).
Assume now that we have a decomposition $\mathcal{H}=\mathcal{H}_1\otimes\mathcal{H}_2$.
Then, the partial traces define a map $\pi:\mathcal{A}\to\mathcal{A}_1\oplus \mathcal{A}_2$.
The evaluation of $\pi$ at a factorizable state  $|v\rangle\langle v| \otimes|w\rangle\langle w|$ gives a pair of pure states
\[
\pi(|v\rangle\langle v| \otimes|w\rangle\langle w|)= 
(|v\rangle\langle v|,|w\rangle\langle w|)\in \mathcal{A}_1\oplus\mathcal{A}_2.
\]
Then, according to the previous definition, the factorizable states (and separable states) are generalized unentangled. In particular, the previous definition generalizes to a more wider situation. It is true that, as remarked by Viola and Barum \cite[p. 18]{ViolaBarnum10}, ``In spite of suggestive points of contact, our approach differs from the [orthodox one] in (at least) two important ways: physically, the need for a decomposition into distinguishable subsystems is bypassed altogether; mathematically, the GE notion rests directly (and solely) on extremality properties of quantum states in convex sets which are associated to different observers.'' However, as we mentioned already, both definitions (i.e., the orthodox and its generalization) are relative to a chosen decomposition $\mathcal{H}=\mathcal{H}_1\otimes\mathcal{H}_2$. As proved in \cite{DelaTorre10}, if we change the decomposition, we arrive at a situation where a (generalized) separable state becomes (generalized) entangled. Finally, let us also remark that it is possible to  give equivalent definitions of generalized entanglement with convex sets instead of cones (see \cite{BKOV05}).

\smallskip

Perspectival-relativism has been always ---since the Ancient Greek sophists--- related to anti-realist trends of thought (see \cite{deRondeFM18}). However, Barnum \cite[p. 348]{Barnum03} seems to propose what we consider to be a very strange idea according to which relativism in QM should be understood as supporting a realist interpretation of the theory: ``I view the relative state interpretation less as a way of getting the classical world to emerge from Hilbert space, and more as a way of giving a realistic interpretation to Hilbert space structure in the presence of additional structures such as preferred bases or subsystem decompositions that represent other aspects of physics.'' In this case, relativism seems to be considered as part of a general realist project which attempts to justify the existence of preferred factorizations and bases. According to this viewpoint, the realist should content herself with finding a way to justify the existence of these preferred perspectives. But while in the case of Earman it is argued that these conditions should come from ``background physics'', Barnum \cite{Barnum90}, advances an even stranger step further ---at lest, for a supposedly realist position--- and argues in favor of considering the role of consciousness: ``As I see it, the correct way, in this view, to account for the appearance that there is a single measurement result is the idea that the experience of a conscious history is associated with definite measurement results, so that consciousness forks when a quantum measurement is made.'' We will come back to these essential points in the following sections. For now we content ourselves to provide an explicit definition of this relativist contextual understanding of entanglement.  

\smallskip 
\smallskip 

\noindent {\it {\sc Contextual Relativism:} Given a pure state, the existence of entanglement is relative to a preferred subset of observables. Thus, the same state will be entangled or not depending on the choice of the specific subset.}

\smallskip 
\smallskip 

To sum up, while the orthodox notion of entanglement presents a definition relative to the factorization of systems, the notion of generalized entanglement replaces this type of relativism by a contextual one. This contextual type of relativism in which observers must choose between the actual existence of different and incompatible states of affairs, all of them contained within the same mathematical representation, is of course highly problematic from a representational realist understanding of physics which understand theories as providing an objective-invariant formal-conceptual representation of a state of affairs ---let it be classical or quantum. But let's keep this on hold. Regardless of these criticisms ---which might be considered by many physicists as merely ``philosophical''--- there are also serious technical problems recognized in the orthodox foundational and informational literature which we would like to address in some detail.

\section{The Many Problems of Orthodox Entanglement}

The exponential growth in the research being done in quantum entanglement due to explosion during the last decades of quantum information processing as one of the most important lines of research worldwide has also lead to the findings of several essential technical difficulties ---and even inconsistencies--- which expose in a different context the serious difficulties which hunt the definitions discussed in the previous sections. Let us mention at least some of them. 

To start with, the presupposed relation between entanglement and {\it non-locality} has been shown  to be unsatisfactory ---to say the least. As resumed by Bokulich and Jaeger: 
\begin{quotation}
\noindent {\small``There are [...] limitations to using a violation of Bell's inequality as a general measure of entanglement. First, there are Bell-type inequalities whose largest violation is given by a non-maximally entangled state (Ac\'in et al. 2002), so entanglement and non-locality do not always vary monotonically. More troublingly, however, Reinhard Werner (1989) showed that there are some mixed states (now referred to as Werner states) that, though entangled, do not violate Bell's inequality at all; that is, there can be entanglement without non-locality. In an interesting twist, Sandu Popescu (1995) has shown that even with these local Werner states one can perform a non-ideal measurement (or series of ideal measurements) that `distills' a non-local entanglement from the initially local state. In yet a further twist, the Horodecki family (1998) subsequently showed that not all entanglement can be distilled in this way ---there are some entangled states that are `bound'. These bound entangled states are ones that satisfy the Bell inequalities (i.e., they are local) and cannot have maximally entangled states violating Bell's inequalities extracted from them by means of local operations. Not only can one have entanglement without non-locality, but also, as Bennett et al. (1999) have shown, one can have a kind of `non-locality without entanglement'. There are systems that exhibit a type of non-local behavior even though entanglement is used neither in the preparation of the states nor in the joint measurement that discriminates the states (see also Niset and Cerf (2006))'' \cite[pp. xvii-xviii]{BokulichJaeger10} }\end{quotation}
As we pointed out earlier, these results are not strange once we recognize that the addressed relation between entanglement and non-locality comes from a widespread misunderstanding regarding the meaning of the violation of Boole-Bell type inequalities in EPR type experiments. Even today, while many physicists still tend to believe that such a violation implies the existence of ``something truly quantum and non-local'', the correct conclusion ---known by the philosophy of QM community--- to be drawn by the violation of Boole-Bell inequalities is that such correlations cannot be represented by a classical theory. Period. The uncritical application of locality in the context of QM is also linked to the notion of separability and its kernel relation to the orthodox definition of entanglement in terms of {\it non-separable states}. Of course,  Einstein himself understood very clearly the difficulties of introducing the principle of separability in QM. As he \cite[p. 172]{Born71} himself remarked in a letter directed to Max Born: ``quantum theory nowhere makes explicit use of this [the separability] requirement.'' Regardless of Einstein's warning, in the contemporary literature {\it separability} has been uncritically translated as (mathematical) {\it factorizability}. This idea is not only ungrounded, it is simply absurd. The factorizability of Hilbert spaces cannot be consistently interpreted as the separability (or dissection) of systems into subsystems. The {\it product} of subspaces cannot be understood as their {\it addition}. The product of two rays (${\cal{H}}_1 \otimes {\cal{H}}_2$) {\it generates} a whole plane (${\cal{H}}_{12} = {\cal{H}}_1 \otimes {\cal{H}}_2$) which obviously contains more points than the ones present when considering the addition of the original rays. While Rob Clifton has discussed in detail the failure of this interpretation in the context of the Schimdt (bi-orthogonal) decomposition \cite{Clifton95, Clifton96}, Dennis Dieks and Andrea Lubberdink have shown that the labeling in the factorization is unrelated to the existence of particles \cite{DieksLubberdink11, Dieks20}: ``These physical labels do not coincide with the factor indices occurring in the total quantum state ---the latter remain associated in the same way with all pure one-particle states even in the classical limit, and therefore cannot refer to individual particles.'' Furthermore, as it is well known in quantum logic due to a famous theorem by Diederik Aerts \cite{Aerts81}, the {\it conjunction} in QM is not the same as the {\it union} of sets and quantum systems must be regarded essentially as non-separable. It is far from clear what is the physical meaning of factorizability in QM and what exactly is its operational content. In fact, as Aerts has argued, one is forced to adapt the mathematical formalism of QM if one is still willing to address separable systems \cite{Aerts84b}. 

The situation is even worse when considering density operators or matrices. It is well known fact that mixed states are a necessary extension beyond pure states for measuring matrices (of rank $\neq$ 1) in the lab. But as explained by Earman \cite{Earman15}: ``Since different convex linear combinations can produce the same mixed state, it is not easy to determine whether or not a given mixed state is indecomposable.'' As also remarked by Thirring et al. \cite{Thirring11}: ``For mixed states, however, the situation is much more complex [than for pure states] (see, e.g., Ref. [17]). The reason is that the maximal mixed state, the tracial state $\frac{1}{D} \mathds{1}_{D}$, is separable for any factorization and therefore a sufficiently small neighborhood D of it is separable too.'' It has becomes a difficult technical problem in the specialized literature to determine weather or not a mixed state is entangled or not \cite{LiQuiao18}. At a more profound level, one will also need to deal with the serious inconsistencies, analyzed in \cite{deRondeMassri19d}, present within the definition of {\it pure state} itself ---also essential in the orthodox definition of entanglement. These inconsistencies also threaten the extension of entanglement to the case of {\it mixtures} \cite{deRondeMassri20a}. 

Also, the orthodox definition of quantum entanglement remains linked to the existence of ``collapses'' which still, almost one century after its {\it ad hoc} introduction, remains with no formal nor experimental support \cite{Gurtovoi20}. As remarked by Dennis Dieks \cite[p. 120]{Dieks10}: ``Collapses constitute a process of evolution that conflicts with the evolution governed by the Schr\"{o}dinger equation. And this raises the question of exactly when during the measurement process such a collapse could take place or, in other words, of when the Schr\"{o}dinger equation is suspended. This question has become very urgent in the last couple of decades, during which sophisticated experiments have clearly demonstrated that in interaction processes on the sub-microscopic, microscopic and mesoscopic scales collapses are never encountered.'' The bottom line of this is an essential one. If collapses do not really exist, then the notion of entanglement, which was created by Einstein and Schr\"odinger in order to target the subjective appearance of collapses in measurement, ---obviously--- needs to be seriously reconsidered. 

Not only the definition, but also the measure of the {\it degree of entanglement} in the lab faces serious technical problems when attempting to follow its orthodox understanding in terms of non-separable pure states. It has been even argued that the degree of entanglement is a physically ill-posed problem \cite{PC06}. Mercin Pawowski and Marek Czachor have analyzed ``an example of a photon in superposition of different modes, and ask what is the degree of their entanglement with vacuum. The problem turns out to be ill-posed since we do not know which representation of the algebra of canonical commutation relations (CCR) to choose for field quantization. Once we make a choice, we can solve the question of entanglement unambiguously.'' The way to characterize and measure the degree of entanglement remains today one of the most serious open problems within the foundational literature. Finally, we might remark that the survey of various operational and non-operational criteria of entanglement by Dagmar Bru\ss \ \cite{Bruss02} points to the serious difficulty that the growing number of competing definitions ---all of them grounded on particle metaphysics and collapses--- threaten to fragment the literature about entanglement in smaller and smaller, compartmented debates with no contact between each other. In fact, for the specific case of indistinguishable particles there is even no consensus regarding the way to define entanglement. As remarked by Rosario Lo Franco and Giuseppe Compagno \cite{LoFrancoCompagno16}: ``Quantum entanglement of identical particles is essential in quantum information theory. Yet, its correct determination remains an open issue hindering the general understanding and exploitation of many-particle systems'' (see also \cite{Dieks20, Earman15}). 

\smallskip 

We might conclude that if no consensus is reached in the field regarding the very basic definition and meaning of quantum entanglement, the literature might end up constructing a tower of Babel in which researchers working in the same field will become unable to understanding each other. Unfortunately, we have already witnessed this fragmentation in understanding within the interpretational debate about quantum theory which has been recently characterized by \'Adan Cabello as ``a map of madness'' \cite{Cabello17}.

\section{An Objective Relational Definition of Quantum Entanglement}

Perspectival-relativism is a special type of co-relationalism which stresses the specific relation between a perceiving subject and its surrounding. Obviously, this particular type of relation goes hand in hand with those positions which assume as a standpoint an anti-realist subject-perspective of analysis ---i..e, subject-object or subject-observation relations. Sophistry, empiricism, positivism, Bohrian philosophy and instrumentalism are all co-relational schemes which take as a standpoint the subjective act of perception given within the process of measurement. As positivists from the Vienna circle justified in their influential manifesto \cite{VC}: ``In science there are no `depths'; there is surface everywhere: all experience forms a complex network, which cannot always be surveyed and, can often be grasped only in parts. Everything is accessible to man; and man is the measure of all things. Here is an affinity with the Sophists, not with the Platonists; with the Epicureans, not with the Pythagoreans; with all those who stand for earthly being and the here and now.'' Starting from a perceptive perspective, observation and measurement become then the necessary foundation of what is considered to be ``scientific''. Observation, considered as an unproblematic {\it given} of ``common sense'' experience, comes always first and physical theories are just mathematical models (with some additional rules) designed to predict future experience. As explained by David Deutsch: 
\begin{quotation}
\noindent {\small``during the twentieth century, most philosophers, and many scientists, took the view that science is incapable of discovering anything about reality. Starting from empiricism, they drew the inevitable conclusion (which would nevertheless have horrified the early empiricists) that science cannot validly do more than predict the outcomes of observations, and that it should never purport to describe the reality that brings those outcomes about. This is known as instrumentalism. It denies that what I have been calling `explanation' can exist at all. It is still very influential. In some fields (such as statistical analysis) the very word `explanation' has come to mean prediction, so that a mathematical formula is said to `explain' a set of experimental data. By `reality' is meant merely the observed data that the formula is supposed to approximate. That leaves no term for assertions about reality itself, except perhaps `useful fiction'.'' \cite[p. 15]{Deutsch04}}
\end{quotation}

Realism, on the other hand, is intrinsically related to the origin of both physics and philosophy, both of which begun with the ancient Greek idea according to which there is something, a fundament, we call {\it physis} (or reality) which comprises the whole of existence. Everything there is, absolutely everything ---including ourselves---, is part of {\it physis}. The second essential  presupposition is that {\it physis} is not chahotic, it has a certain order or what the Greeks called a {\it logos}. Such a {\it logos} is expressible through the creation of {\it theories}. It is a difficult task to expose the true {\it logos} since, as remarked by Heraclitus, ``{\it physis} loves to hide.'' [f. 123 DK]. Doing so requires hard work and sensibility, but ---following Heraclitus--- the latter can be revealed in the former. In a particular {\it logos} one can ``listen'' something that exceeds it, that is not only that personal discourse but the logos of {\it physis}: ``Listening not to me but to the {\it logos} it is wise to agree that all things are one'' [f. 50 DK]. We are thus able to represent {\it physis}, to exhibit its {\it logos}.\footnote{As shown by Heraclitus, the logos of theories was never supposed to mirror the logos of {\it physis}. One allows to reach the other. However, the idea that realism implies a a one-to-one {\it correspondence relation} between theory and reality has been systematically applied by anti-relaists in order to create a straw-realist they can fight and easily destroy.} By being able to subsume the many different phenomena within a theoretical unity, the claim made by the first realists was that theories were able to capture or express, in some way, aspects of reality. As we all know, this idea was strongly confronted by sophists, giving rise to an endless war between realists and anti-relaists. 

Thus, we might regard the main point of disagreement between realists and anti-realists as a mater of perspective. As Thomas Nagel described, the anti-realist is a perspective from {\it now here} in contraposition to the realist which he ironically referred to as a perspective from {\it no where} (see also {\cite{Barnum03, Sudbery16}). While the first group assumed the down to earth perspective of a perceiving subject, the latter assumed the theoretical perspective of {\it physis}. This makes all the difference. For realists, theoretical representation comes always before observation. The conditions of observation are derived conceptually and formally from a theory, not the other way around. As Einstein famously said to Heisenberg: ``it is only the theory which can tell you what can be observed.'' Only adequate concepts are capable to explain what has been observed. As Heisenberg \cite[p. 264]{Heis73} himself would make the point: ``For an understanding of the phenomena the first condition is the introduction of adequate concepts. Only with the help of correct concepts can we really know what has been observed.'' From a realist perspective, observation can never affect the representation of a state of affairs. As remarked by Einstein \cite[p. 175]{Dieks88a}: ``[...] it is the purpose of theoretical physics to achieve understanding of physical reality which exists independently of the observer, and for which the distinction between `direct observable' and `not directly observable' has no ontological significance.'' Observation and measurement are nonstarters for any true realist. Realism is not a subjective belief in an interpretation, it is a {\it praxis}, a specific  procedure in which physicists seek to create invariant-objective (subject-independent) theoretical (formal-conceptual) representations of states of affairs (see for a more detailed discussion \cite{deRonde20b, deRonde20c}). Examples of such theoretical productions are Newtonian classical mechanics, Maxwell's theory of electromagnetism or Einstein's Relativity. The program implies the need to produce global formal-conceptual representations in which all reference frames and observational perspectives are consistently considered allowing in this way to detach the subject from the represented course of events.

The specific manner in which theories are able to create (subject-independent) representations of states of affairs is intrinsically related to two kernel notions of modern physics, namely, objectivity and operational-invariance. While the first is one of a conceptual nature, the latter is one of a mathematical or formal character. One is the counterpart of the other. Both act as preconditions of consistency regarding the global representations constructed by the realist. Of course, within such representations there is no room for preferred reference frames (bases) nor preferred observational or experimental perspectives. Let us discuss this in some more detail. {\it Objectivity} means the provision of a categorical scheme which is able to account for a multiplicity of phenomena in terms of a conceptual {\it moment of unity}. Following Kant, the notion of `object' must be regarded as a conceptual machinery capable not only of creating a representational realm in which thinking of experience becomes possible but also qualified to unify in a consistent manner the different phenomena observed by empirical subjects. In physics, all perspectives must be considered as equally acceptable. This is in fact the democratic pre-condition of scientific objectivity.  Realism imposes the need that all representations from different viewpoints must be subsumed within the consistent account of the same state of affairs. Even though there might exist a multiplicity of different perspectives of analysis, reality must be considered as one and the same ---independent of any particular perspective. {\it Operational-invariance} in physical theories capture exactly this requirement in mathematical terms. In this case, the mathematical representation of the {\it state} of a system (i.e., the object in a specific situation) is given from a specific reference frame (or basis) which can be consistently translated into any other reference frame (or basis) (see for a detailed analysis \cite{deRondeMassri17, deRondeMassri18a, Scholz18}). While in the case of classical mechanics this translation is provided  by the Galilean transformations, in the case of relativity theory this is given through the Lorentz transformations. This is what defines the realist program: the attempt to produce a consistent, coherent and unified theory, which is also the attempt to produce an invariant-objective (formal-conceptual) representation of both reality and experience. 

\smallskip

In the particular case of QM, the possibility of such a realist development has been investigated by the logos categorical approach. An approach which, avoiding metaphysical prejudices, attempts to follow the realist procedure of theory construction. According to this procedure, since in QM we already have an operational mathematical formalism capable to operationally account for experience, the focus must be centered in the development of a new conceptual scheme which is able to provide a consistent account of the mathematical formalism. We already know that the classical atomist representation does not work for this mathematical scheme; thus, we need to produce a set of completely new non-classical concepts which match the formalism in an adequate manner ---and not the other way around. The thread of Ariadna which has allowed us to escape the Bohrian-positivist labyrinth has been the well known operational-invariance of the theory of quanta itself, namely, the Born rule. It is this mathematical expression of the theory which captures the essential connection between the theory, particular reference frames (or bases) and experience. The Born rule provides the operational invariant content of QM which, leaving behind dogmatic (atomist) metaphysics and certain (binary) observations, can be related to a new understanding and generalization of EPR's famous element of physical reality.  According to it, the probabilistic information supplied by the Born rule should be considered as providing objective knowledge of a state of affairs \cite{deRonde16} ---instead of the probabilistic prediction of a single measurement outcome. This can be done given we abandon the {\it binary} reference to actual properties and certain observations and advance into an {\it intensive} representation of a state of affairs. Escaping relativism, we have proven in \cite{deRondeMassri18a} that our scheme is capable of bypassing the Kochen-Specker theorem through the introduction of a {\it global intensive valuation} which, in turn, allows us also to produce an objective (subject-independent) invariant (basis-independent) account of the orthodox  mathematical formalism. Let us see this in some more detail. 

Let us begin by recalling some results from \cite{deRondeMassri18a} which allow us to  introduce some essential concepts. A \emph{Global Intensive Valuation} (GIV) is a function from a graph to the closed interval $[0,1]$. 
A \emph{Global Binary Valuation} (GBV) is a function from a graph to 
the set $\{0,1\}$.  The graph that we are interested in, is the graph of projection operators which we term intensive powers. Let $\mathcal{H}$ be a Hilbert space and let $\mathcal{G}=\mathcal{G}(\mathcal{H})$  be the set of observables. We give to $\mathcal{G}$ a graph structure by assigning an edge between observables $P$ and $Q$ if and only if $[P,Q]=0$. We call this graph, \emph{the graph of powers}. Among all global intensive valuations we are interested in the particular class of PSA.
\begin{definition}
Let $\mathcal{H}$ be a Hilbert space.
A \emph{Potential State of Affairs} is a global intensive valuation
$\Psi:\mathcal{G}(\mathcal{H})\to[0,1]$ from the graph of powers $\mathcal{G}(\mathcal{H})$
such that $\Psi(I)=1$ and 
\[
\Psi(\sum_{i=1}^{\infty} P_i)=
\sum_{i=1}^\infty \Psi(P_i)\]
for any piecewise orthogonal projections $\{P_i\}_{i=1}^{\infty}$.
The numbers $\Psi(P) \in [0,1]$, are called {\it intensities} or {\it potentia}
and the nodes $P$ are called \emph{ powers}.
Hence, a PSA assigns a potentia to each power.
\end{definition}
Intuitively, we can picture a PSA
as a table,
\[
\Psi:\mathcal{G}(\mathcal{H})\rightarrow[0,1],\quad
\Psi:
\left\{
\begin{array}{rcl}
P_1 &\rightarrow &p_1\\
P_2 &\rightarrow &p_2\\
P_3 &\rightarrow &p_3\\
  &\vdots&
\end{array}
\right.
\]

\begin{theorem}
Let $\mathcal{H}$ be a separable Hilbert space, $\dim(\mathcal{H})>2$ and let $\mathcal{G}$ be the graph of immanent powers with the commuting relation given by QM.
\begin{itemize}
\item Any positive semi-definite self-adjoint operator 
of the trace class $\rho$ determines in a bijective way
a PSA $\Psi:\mathcal{G}\to [0,1]$. 
\item Any GIV determines univocally a GBV such that the set of powers are considered as potentially existent. 
\end{itemize}
\end{theorem}
\Proof
See \cite{deRondeMassri18a}.
\cqd

\begin{definition}
Let $\mathcal{G}$ be a graph. We define a \emph{context} as a complete subgraph (or aggregate) inside $\mathcal{G}$. For example, let $P_1,P_2$ be two elements of $\mathcal{G}$. Then, 
$\{P_1, P_2\}$ is a contexts if $P_1$ is related to $P_2$, $P_1\sim P_2$. Saying it differently, if there exists an edge between $P_1$ and $P_2$. In general, a collection of elements $\{P_i\}_{i\in I}\subseteq \mathcal{G}$ determine a context if $P_i\sim P_j$ for all $i,j\in I$. Equivalently, if the subgraph with nodes $\{P_i\}_{i\in I}$ is complete.  A \emph{maximal} context is a context not contained properly in another context.  If we do not indicate the opposite, when we refer to contexts we will be implying maximal contexts.
\end{definition}

For the graph of powers, the notion of context coincides with the usual one; a complete set of commuting operators. However, all projection operators can be assigned a consistent value bypassing in this way the famous Kochen-Specker theorem.

\begin{theorem} 
{\sc (Intensive Non-Contextuality Theorem)} If $\mathcal{H}$ is a Hilbert space, then a 
PSA is possible.
\end{theorem} 
\Proof
See \cite{deRondeMassri18a}.
\cqd

\smallskip

\noindent This theorem restores the possibility of an objective physical representation of any quantum wave function $\Psi$. Contrary to the orthodox interpretation of QM in terms of systems with properties which imply a binary valuation, our conceptual representation of quantum physical reality is not relative to any particular context, it is global and essentially intensive.

\smallskip

In turn, our logos approach has allowed us to reconsider the notion of entanglement from a purely mathematical standpoint in terms of the actual and potential coding of effective and intensive relations between different screens \cite{deRondeMassri19b}. Let us see this in some detail. 

\medskip

\noindent {\it
{\sc Effective Relations:} The relations determined by a difference of possible actual effectuations. Effective relations discuss the possibility of an actualist definite coding. It involves the path from intensive relations to definite correlated (or anti-correlated) outcomes. They are determined by a binary valuation of the factorized context in which only one correlated node is considered as true in each subgraph, while the rest are considered as false.} 

\medskip

\noindent {\it
{\sc Intensive Relations:} The relations determined by the intensity of different powers. Intensive relations imply the possibility of a potential intensive coding. They are determined by the correlation of intensive valuations.}

\medskip 

\noindent These relations provide an intuitive grasp of what can be done in a lab and what type of relations are at play. The following definitions provide a new account of entanglement which rests on the analysis of relational intensive and effective correlations.
\begin{dfn}[Quantum Entanglement]
Given $\Psi_1$ and $\Psi_2$ two PSAs, if $\Psi_1$ and $\Psi_2$ are related intensively and effectively we say there exists \emph{quantum entanglement} between $\Psi_1$ and $\Psi_2$.
\end{dfn}
According to this definition entanglement relates to the potential coding of intensive and effective relations between two distant measuring set-ups. We also have the possibility to provide an intuitive non-spatial definition of separability which relates to the lack of correlations between two distant screens.
\begin{dfn}[Relational Separability]
Given $\Psi_1$ and $\Psi_2$ two PSAs, if $\Psi_1$ and $\Psi_2$ are not related intensively nor effectively we say there is \emph{relational separability} between $\Psi_1$ and $\Psi_2$.
\end{dfn}
It is interesting to notice that our definitions of potential coding in terms of intensive and effective relations allows us to address a third possibility which considers the cases in which there are only intensive relations involved but not effective ones.
\begin{dfn}[Intensive Correlation]
Given $\Psi_1$ and $\Psi_2$ two PSAs, if $\Psi_1$ and $\Psi_2$ are related intensively but not effectively we say there exists an \emph{intensive relation} between $\Psi_1$ and $\Psi_2$.
\end{dfn}

This new approach shows not only that it is possible to derive an objective representation for the theory of quanta but also that metaphysical considerations are essential for the analysis of operational data. As we have discussed in detail in \cite{deRondeFreytesSergioli19}, the reference to intensive data provides a completely new perspective of analysis for quantum information processing. The essential relevance of intensive relations in QM has been completely bypassed within the orthodox definition of entanglement which has been exclusively focused on the dogmatic belief in the existence of invisible and irrepresentable particles which collapse when observed \cite{deRondeMassri20b}. This dogmatic beliefs have acted in the philosophical and foundational communities as epistemological obstructions which have boycotted the possibility of thinking about entanglement in new original and consistent ways. It must be clear that since the orthodox notion of entanglement is essentially grounded on both `particle metaphysics' and the existence of `collapses', the rejection of these concepts also implies the rejection of the present definition of quantum entanglement. Our redefinition beyond particles and {\it ad hoc} rules hopes to open a necessary debate about the notion of quantum entanglement.

In the following section we attempt to show that the logos definition of entanglement is capable to bypass both the non-separable relativism (section 2) and the contextual relativism (section 3) which hunt the orthodox definitions.

\section{Relational Quantum Entanglement Beyond Relativism}

During the 1980s the Finish physicist Kalervo Laurikainen focused his research in philosophical issues about QM and, more specifically, the thought of Wolfgang Pauli, a key figure standing just in between Niels Bohr and Albert Einstein. According to Laurikainen: 
\begin{quotation}
\noindent {\small ``It is not generally known that there was a profound
difference in the philosophical attitudes of Niels Bohr and Wolfgang
Pauli (Laurikainen 1985b, section 3). In his address at the {\it Second
Centenary of Columbia University} in 1954, `The Unity of
Knowledge', Bohr claimed that the observer even in quantum
mechanics can be considered `detached' provided we understand the
observation in the right way (Bohr 1955, p. 83). An observation
includes a detailed description of all the experimental arrangements
which can have an influence upon the phenomenon under investigation,
and it is finished only when a registered result is obtained which
everybody can verify afterwards. In this sense, Bohr said, an
observation is quite {\it {\small objective}} (which for Bohr means
`intersubjective'), and the observer does not have any influence on
the result in any other way than by choosing the method of
observation. The result is explicitly associated with a given method
of observation. If physics is understood as a system which makes it
possible to govern such objective observational results ---which,
however, is only possible in probabilistic sense--- then physics,
according to Bohr, can even in atomic physics be considered quite
objective and the observer is `detached' in exactly the same way as
in classical physics.'' \cite[p. 42]{Laurikainen98}} \end{quotation}
Bohr had sent the manuscript of the paper to Pauli in order to
receive his critics and comments. In his reply, dated February 15,
1955, Pauli tried to explain him ---in line with Einstein--- that
the role of the observer in classical mechanics was essentially
different from the one presented in the orthodox collapse interpretation of QM.
\begin{quotation}
\noindent {\small ``[...] it seems to me quite appropriate to call the
conceptual description of nature in classical physics, which
Einstein so emphatically wishes to retain, `the ideal of the
detached observer'. To put it drastically the observer has according
to this ideal to disappear entirely in a discrete manner as hidden
spectator, never as actor, nature being left alone in a
predetermined course of events, independent of the way in which
phenomena are observed. `Like the moon has a definite position'
Einstein said to me last winter, `whether or not we look at the
moon, the same must also hold for the atomic objects, as there is no
sharp distinction possible between these and macroscopic objects.
Observation cannot {\it create} an element of reality like position,
there must be something contained in the complete description of
physical reality which corresponds to the {\it possibility} of
observing a position, already before the observation has been
actually made.' I hope, that I quoted Einstein correctly; it is
always difficult to quote somebody out of memory with whom one does
not agree. It is precisely this kind of postulate which I call the
ideal of the detached observer.

In quantum mechanics, on the contrary, an observation {\it hic et
nunc} changes in general the `state' of the observed system, in a
way not contained in the mathematical formulated {\it laws}, which
only apply to the automatical  time dependence of the state of a
{\it closed} system. I think here of the passage to a new phenomenon
of observation which is taken into account by the so-called
`reduction of the wave packets'. As it is allowed to consider the
instruments of observation as a kind of prolongation of the sense
organs of the observer, I consider the impredictable change of the
state by a single observation ---in spite of the objective character
of the results of every observation and notwithstanding the
statistical laws of frequencies of repeated observation under equal
conditions--- to be \emph{an abandonment of the idea of the
isolation (detachment) of the observer from the course of physical
events outside himself.} 

To put it in nontechnical common language one can compare the role
of the observer in quantum theory with that of a person, who by his
freely chosen experimental arrangements and recordings brings forth
a considerable `trouble' in nature, without being able to influence
its unpredictable outcome and results which afterwards can be
objectively checked by everyone.'' \cite[p. 60]{Laurikainen88}}
\end{quotation}
Bohr's misunderstanding with respect to the meaning of objectivity went hand in hand with his misunderstanding of the fact that Einstein's relativity theory did not imply any sort of perspectival-relativism. Just like Bohr had confused the notion of {\it intersubjectivity} with that of {\it objectivity}, Bohr also confused in several occasions his relativist-perspectival interpretation of QM with Einstein's relativity theory. In his Commo paper from 1929 he wrote the following: 
\begin{quotation}
\noindent {\small ``While the theory of relativity reminds us of the subjective character of all physical phenomena, a character which depends essentially upon the state of motion of the observer, so does the linkage of the atomic phenomena and their observation, elucidated by the quantum theory, compel us to exercise a caution in the use of our means of expression similar to that necessity in psychological problems where we continually come upon the difficulty of demarcating the objective content.'' \cite[p. 116]{Bohr34}}
\end{quotation} 
This fragment shows the deep failure of Bohr not only to understand the physics behind Einstein's theory of relativity, but more importantly, the role of objectivity and invariance in physics as the main conditions for a ``detached subject'' representation ---as contra-posed to his own perspectival-relativist approach.\footnote{This essential mistake is also present in Bohr's famous reply to EPR \cite[p. 702]{Bohr35}: ``The dependence on the reference system, in relativity theory, of all readings of scales and clocks may even be compared with the essentially uncontrollable exchange of momentum or energy between the objects of measurements and all instruments defining the space-time system of reference, which in quantum theory confronts us with the situation characterized by the notion of complementarity. In fact this new feature of natural philosophy means a radical revision of our attitude as regards physical reality, which may be paralleled with the fundamental modification of all ideas regarding the absolute character of physical phenomena brought about by the general theory of relativity.''} 

Unfortunately, this confusion has become completely widespread in many relativist-perspectival interpretations of QM which regardless of their impossibility to provide a global representation of the theory still claim ---following Bohr's rhetorics--- to be both realist and objective \cite{deRondeFM18}. Let us try to be very clear about this essential point. It is obviously true that physical theories require in order to account for any given situation the specification of a {\it reference frame}. This is maybe the very first lesson that one must learn as a physicist, namely, that the representation of a state of affairs always requires the explicit specification of the formal perspective from which the situation is being described. Even in classical mechanics, the position or velocity of a body have meaning only with respect (i.e., relative) to a frame of reference.  But of course, this relativism with respect to reference frames has no ontological content. This relativism evaporates immediately when considering the group of transformations which allows us to shift the representation of the state of affairs from one reference frame to another in a consistent (invariant) manner.\footnote{As shown in \cite{deRondeMassri17}, this consistency in the translation of values of properties for different reference frames is what perspectival interpretations of QM lack completely. Perspectival or relativist interpretations of QM have no global representational unity, and that is the reason measurement set-ups and preferred bases become essential.} And it is exactly this same operational-invariance ---of the values of the properties relative to different reference frames--- which allows us to {\it detach} all reference frames (or empirical subjects) from the particular representation of a (real) state of affairs. In fact, the whole point of talking about ``something real'', is to talk about something independent of any preferred (formal or empirical) perspective. This is the essentially democratic nature of science which goes against any one claiming to have a preferred access to reality. The realist cannot accept the existence of a {\it preferred basis} or a {\it preferred context} in order to consider the representation of a state of affairs. 

Going back to our quotation, the analogy that Bohr attempted to make between relativity and QM is obviously wrong: Relativity theory is in no way different from classical mechanics with respect to invariance and objectivity. None of them imply anything like the existence of the ``subjective character of all physical phenomena''. On the contrary, both theories are objective and operationally invariant allowing a detached subject representation of any state of affairs. The only difference  between them is that while in classical mechanics it is the Galilean transformations which allows us to consider all reference frames as consistently referring to the same real state of affairs, in the case of relativity theory this is done through the Lorentz transformations. As Max Jammer \cite[p. 132]{Jammer74} remarked about our just quoted passage: ``Bohr overlooked that the theory of relativity is also a theory of invariants and that, above all, its notion of `events,' such as the collision of two particles, denotes something absolute, entirely independent of the reference frame of the observer and hence logically prior to the assignment of metrical attributes.'' The only thing that is ``subjective'' here is Bohr's own interpretation of QM, where there is no global consistency of the (binary) values of projection operators considered from different contexts.\footnote{The impossibility of a global binary valuation was later on explicitly proved by the now famous Kochen-Specker theorem. See for a detailed analysis \cite{deRonde20a}.} As Bohr famously argued in his reply to the EPR paper, the condition to consider a subset of observables as definite valued was the specification of the reference frame (or basis). Thus, in Bohr's approach ---unlike in Einstein's relativity theory--- the description of a state of affairs would become intrinsically {\it relative} to the agent's choice of a single (preferred) context (or basis). Of course, this goes in line with Bohr's claim that in the great drama of quantum existence the subject should be regarded not only a spectator but also as actor ---something untenable in the theory of relativity. 

As we have shown explicitly in \cite{deRondeMassri17, deRondeMassri18a} the inconsistency regarding the value invariance of properties is not a consequence of the mathematical formalism of QM, it is only specific to the consideration of {\it binary valuation} imposed by a dogmatically presupposed actualist (metaphysical) understanding of physics. As we mentioned above, the failure to understand the consequences of relativism seem also present in Barnum's considerations \cite[p. 348]{Barnum03} who restricts the role of realism ---like Earman also does--- to the mere justification of the existence of preferred bases and factorizations: ``An important point brought out by the attempt at a relative state interpretation of quantum mechanics is the need to bring in, in addition to Hilbert space, notions of preferred subsystems (`experimenter' and `system' perhaps also the `rest of the world') or preferred orthogonal subspace decompositions (choice of `pointer basis' (Zurek, 1981)).'' As we have attempted to make clear, one of the essential points of being a realist is to reject any preferential perspective within the representation of a state of affairs. It is exactly this independence from preferred viewpoints, reference frames or basis, which has been accomplished in the logos formulation of QM through the consideration of a {\it Global Intensive Valuation} \cite{deRondeMassri18a}. If the notion of entanglement is to be understood as a feature of the real word, and not merely a way in which subjects make choices about fictitious entities, then we must be able to provide an objective account of its relations independently of the choice of reference frames (bases) or mathematical decompositions (factorizations). In the following, we derive two theorems which show explicitly how the logos definition of entanglement is capable to escape the relativism that hunts the orthodox definitions.  

\smallskip 

\smallskip

By reinterpreting a result from \cite{deRondeMassri18a} and using our previous considerations we can derive the following important theorem.
\begin{theorem}{\sc (Basis Invariance Theorem)}
Let $\mathcal{H}$ be a Hilbert space, $\mathcal{G}$ its graph of powers and let 
$\Psi:\mathcal{G}\to[0,1]$ be a PSA. Let $\mathcal{C}_1,\mathcal{C}_2\subseteq\mathcal{G}$ be two contexts. The intensities of $\Psi$ over $\mathcal{C}_1$ and $\mathcal{C}_2$ may be different, but both determine in a unique way the whole PSA. In other words, the GIV defined by $\Psi$ does not depend on the basis for $\mathcal{H}$. In particular, since effective and intensive relations are defined globally, the definitions of quantum entanglement, relational separability and intensive correlation are non-contextual.
\end{theorem}
As a consequence of the {\it Basis Invariance Theorem} just stated, the relational definition of entanglement is shown to be independent of the choice of any particular basis or context. The choice of the basis or context becomes in this scheme just the choice of a viewpoint of analysis which is compatible with the choice of any other viewpoint. 

\smallskip 

\smallskip 

Let us now analyze the factorization problem posed in \cite{DelaTorre10}.
Let us denote by ${\cal PSA}(H)$ the set of all possible PSAs on 
the Hilbert space $H$. We can identify this set (after fixing a basis)
with the space of density matrices over $H$.
In fact, given any completely positive trace-preserving 
map $T:B(H_1)\to B(H_2)$, we obtain a map $T_{*}:{\cal PSA}(H_1)\to {\cal PSA}(H_2)$.
An example of a completely positive trace-preserving map is the partial trace.
Notice that for each factorization $H=H_1\otimes H_2$, we have 
a different partial trace, $T:B(H)\to B(H_1)$.
\begin{definition}
Let $T:B(H)\to B(H')$ be a
completely positive trace-preserving map between (trace-class) operators on Hilbert spaces $H,H'$ and
let $\Psi:\mathcal{G}(H)\to[0,1]$ be a PSA over $H$.
We say that $\Psi':\mathcal{G}(H')\to[0,1]$ is a \emph{shadow}
of $\Psi$ if $\Psi'=T_*(\Psi)$.
Specifically, if $\Psi$ is given by a density matrix $\rho$. 
Then, $\Psi'$ is given by $\rho'=T(\rho)$.
\end{definition}
The relevance of this definition is that it allows us to avoid the factorization problem posed by De la Torre et al.
\begin{theorem} {\sc (Factorization Invariance Theorem)}
Let $\mathcal{H}$ be a Hilbert space with factorizations $\mathcal{H}=\mathcal{H}_1\otimes\mathcal{H}_2=
\mathcal{H}_1'\otimes \mathcal{H}_2'$. Let $T:B(\mathcal{H})\to B(\mathcal{H}_1)$ and $T':B(\mathcal{H})\to B(\mathcal{H}_1')$
be the partial traces of these factorizations.
Let $\Psi$ be a PSA over $\mathcal{H}$ and let $\Psi_1=T_*(\Psi)$ and $\Psi_1'=T_*'(\Psi)$ the corresponding PSAs over $\mathcal{H}_1$
and $\mathcal{H}_1'$. 
Assume there exists a completely positive trace-preserving map 
$U:B(\mathcal{H}_1)\to B(\mathcal{H}_1')$
such that $UT=T'$, or diagrammatically,
\[
\xymatrix{
&B(\mathcal{H})\ar[ld]_{T}\ar[rd]^{T'}&\\
B(\mathcal{H}_1)\ar[rr]^U&&B(\mathcal{H}_1')
}
\]
Then, $\Psi_1'=U_* (\Psi_1)$.
\end{theorem}
\Proof
Straightforward, $\Psi_1'=T'_*(\Psi)=(UT)_*(\Psi)=U_* T_* (\Psi)=U_*(\Psi_1)$.
\cqd

The previous theorem implies that all factorizations are consistent and that all corresponding {\it shadows} of the PSA are all compatible between each other. In fact, if we have a compatibility $U$ between the factors $\mathcal{H}_1$ and $\mathcal{H}_1'$, then this compatibility translates itself in a compatibility between $\Psi_1$ and $\Psi_1'$.
\begin{corollary}
All factorizations are compatible with respect to the same PSA.
\end{corollary}
\Proof
Follows from the previous Theorem.
\cqd
\begin{remark}
Notice that there is a deep relation between factorizing a Hilbert space as $\mathcal{H}=\mathcal{H}_1\otimes\mathcal{H}_2$ and choosing a basis for $\mathcal{H}$. In fact, given a basis $\{v_i\}$ for $\mathcal{H}_1$ and $\{w_j\}$ for $\mathcal{H}_2$, we can construct a basis for $\mathcal{H}$ as $\{v_i\otimes w_j\}$. Then, we can say that
a factorization determines a basis for $\mathcal{H}$ from the basis of its factors.
\end{remark}
Thus, the relational definition of entanglement is immune to the contextual relativism discussed in \cite{DelaTorre10}. The choice of the factorization has no incidence in the intensive and effective relations already contained in the considered PSA. In the logos approach the choice of the factorization makes reference to the number of intensive powers considered within a PSA. All choices of different factorizations remain consistent with the same global account of the PSA.

\smallskip 

\smallskip 

To sum up, we might conclude that the non-separable and contextual relativisms found in the orthodox interpretations is bypassed in the new objective definition of entanglement. The choice of bases and factorizations does not change the intensive and effective relations already contained within the given (potential) state of affairs described by QM in terms of powers with definite potentia. Thus, all powers and potentia in a given (potential) state of affairs can be considered as objectively existent, invariant and independent of any particular measurement.

\section{Final Remarks}

Erwin Schr\"odinger wrote in 1935 that quantum entanglement should be regarded as ``the characteristic trait of quantum mechanics, the one that enforces its entire departure from classical lines of thought.'' It took more than fifty years for the community of physics to acknowledge this fact. As John Earman \cite{Earman15} makes the point: ``With the rise of quantum computing and quantum information theory there has been a sea change in attitude towards entanglement: it is not something to be feared but rather is a resource to be exploited.'' However, as pointed out by Alisa Bokulich and Greeg Jaeger \cite[p. xiv]{BokulichJaeger10}: ``Despite the fact that the phenomenon of entanglement was recognized very early on in the development of quantum mechanics, it remains one of the least understood aspects of quantum theory.'' We believe that the deep misunderstandings surrounding this kernel notion in the orthodox literature are intrinsically related, in part to the widespread instrumentalist approach of physics and in part to the dogmatic reference to atomist metaphysics. Both instrumentalism and metaphysical atomism act today as a epistemological obstacles which preclude the possibility of a truly (non-classical) conceptual approach to the theory of quanta. In this same respect, the rise of entanglement in the new millennia might be taken as a lesson of the essential role of philosophical conceptual analysis within physics. It is our conviction that it will not be possible to continue this revolution without truly comprehending the meaning of entanglement. Such a comprehension implies the need to go not only beyond the instrumental operational prediction of `clicks' in detectors but also beyond the misleading discourse about ``small particles'' living in an unobservable microscopic world.




\section*{Acknowledgements} 

Christian de Ronde wants to thank Diederik Aerts and D\'ecio Krause for discussions on related subjects. This work was partially supported by the following grants: FWO project G.0405.08 and FWO-research community W0.030.06. CONICET RES. 4541-12 and the Project PIO-CONICET-UNAJ (15520150100008CO) ``Quantum Superpositions in Quantum Information Processing''.


\end{document}